\documentclass[
prl,reprint,
superscriptaddress,
showpacs,
amsmath,amssymb,
aps
]{revtex4-1}

\usepackage{graphicx}
\usepackage{amsmath}
\usepackage[usenames]{color}
\setcitestyle{super}

\definecolor{newcolor}{rgb}{0.9,0,0.1}

\newcommand{\figref}[1]{Fig.~\ref{#1}}
\newcommand{\unit}[2]{$#1\,\text{#2}$}
\newcommand{\equnit}[3]{$\text{#1}=#2\,\text{#3}$}

\begin{document}

\title{Effect of electron--phonon interaction on the formation of one-dimensional electronic states in coupled Cl vacancies}

\author{Bruno Schuler}
\email[]{bsc@zurich.ibm.com}
\affiliation{IBM Research -- Zurich, S\"aumerstrasse 4, 8803 R\"uschlikon, Switzerland}

\author{Mats Persson}
\affiliation{Surface Science Research Centre, Department of Chemistry, University of Liverpool, Liverpool, L69 3BX, United Kingdom}
\affiliation{Department of Applied Physics, Chalmers University of Technology, SE 41296, G\"oteborg, Sweden}

\author{Sami Paavilainen}
\affiliation{Department of Physics, Tampere University of Technology, 33720 Tampere, Finland}

\author{Niko Pavli\v{c}ek}
\author{Leo Gross}
\author{Gerhard Meyer}
\affiliation{IBM Research -- Zurich, S\"aumerstrasse 4, 8803 R\"uschlikon, Switzerland}

\author{Jascha Repp}
\affiliation{Institute of Experimental and Applied Physics, University of Regensburg, 93053 Regensburg, Germany}

\date{\today}


\begin{abstract}
{\bf The formation of extended electron states in one-dimensional
  nanostructures is of key importance for the function of molecular
  electronics devices. Here we study the effects of strong
  electron--phonon interaction on the formation of extended electronic
  states in intentionally created Cl vacancy pairs and chains in a
  NaCl bilayer on Cu(111). The interaction between the vacancies was
  tailored by fabricating vacancy pairs and chains of different
  orientation and separation with atomic precision using vertical
  manipulation. Small separation of divacancies led to the formation
  of symmetric and antisymmetric vacancy states and localized
  interface-states. By scanning tunneling spectroscopy (STS) we
  measured their energy splitting and broadening as a function of the
  inter-vacancy separation. Unexpectedly, the energy splitting between
  the vacancy states is enlarged by level repulsion resulting from
  phonon dressing of the electronic states, as evidenced by
  theory. Already for a few coupled vacancies we observe an emerging
  band structure of the defect band.}
\end{abstract}

\maketitle

One-dimensional nanostructures assembled from single metal atoms and
molecules on surfaces are of great fundamental and technological
interest as their electronic states exhibit a wealth of new quantum
phenomena that might be exploited in future molecular electronics devices~\cite{oncel2008}. Some examples of such phenomena identified by
scanning tunneling microscopy (STM) and spectroscopy (STS) include
metal-atom chains exhibiting 'particle-in-box'
states~\cite{Nilius2002}, bistable antiferromagnetism~\cite{Loth2012},
Tomonaga--Luttinger liquid behaviour~\cite{Bluetal2011}, and signs of
Majorana spinors~\cite{Nadj-Perge2014}. The formation of extended or
band-like electronic states in metal-atom and molecular chains on
surfaces has been widely studied experimentally~\cite{Crain2005,Barke2007,Folsch2009,Wang2011,Riss2014}, but
potentially interesting effects of strong electron--phonon coupling on
these states, such as, for instance, polaronic effects, have attracted
much less attention. A notable exception is the observation of
coherent electron--phonon states and the breakdown of the
Born--Oppenheimer approximation in oligothiophene molecular
chains~\cite{Repp2010}. In the electronically adiabatic regime, where
the Franck--Condon principle is applicable, the electron--phonon coupling
results in vibronic satellite structures. How these vibronic effects
influence the delocalization of an injected electron in a
one-dimensional nanostructure is an open question, in particular when
the electron--phonon coupling is so strong that these effects
dominate. \\

Here we address this question in a combined experimental and
theoretical study of coupled electronic states localized at
intentionally created Cl divacancies and vacancy chains in a NaCl
bilayer on Cu(111). The Cl vacancies provide a quantum-well structure
that could be created with atomic precision using the STM tip~\cite{Olsson2004,folsch2014}. The
energies of the electronic states formed in these structures by
injection of an electron from the tip of an STM and their spatial
behaviour on the atomic scale were resolved by STS and analyzed using simple models. Our findings have
ramifications for the formation of extended states from coupled
adsorbate or confined defect states (such as in
dopants~\cite{koenraad2011} or dangling bonds~\cite{schofield2013}) in semiconductors and ionic crystals.
Hence, it could improve the understanding of polaronic effects in
coupled quantum systems, such as quantum
dots~\cite{lorke1990coupling,Hameau1999,haider2009controlled,Sun2009,leturcq2009,Seufert2013}, charge-based qubits~\cite{andresen2007charge,Mujica2013} and quantum cellular automata~\cite{lent1993quantum}.\\

Coupled, individual Cl vacancies in a NaCl bilayer on a Cu(111)
surface are ideal model systems to study the effect of strong
electron--phonon coupling on the delocalization of electronic
states. First, Cl monovacancies are atomically well-defined and highly
symmetric defects. They feature two distinct electronic states: an
unoccupied vacancy state (VS) that is strongly coupled to optical
phonons in the film and a localized interface-state (LIS) with a
negligible electron--phonon
coupling~\cite{nilius2003influence,Repp2005}. The electronic coupling
between VSs or LISs of different vacancies can be tuned in a
controlled way by the lateral spacing between vacancies with atomic
precision~\cite{nilius2003distance}. Finally, Cl vacancies feature a
much higher stability against inelastic excitations than
adsorbates~\cite{sonnleitner2011}. Cl vacancies could even be filled with other atoms~\cite{kawai2014}. 
Hence, they allow us to explore
artificial coupled quantum systems with great control.\\

In a previous study~\cite{Repp2005a}, single Cl vacancies in a NaCl
bilayer on copper surfaces were characterized in detail. A Cl vacancy
in the outermost layer of this supported NaCl bilayer is analogous to
the widely studied and well-known colour centers in bulk
NaCl~\cite{Zielasek2000} with one key difference: the localized VS is
unoccupied because the electron can tunnel into unoccupied metal
states. The unoccupied VS then gives rise to a positively charged
vacancy and an attractive potential that is able to split off a LIS
from the free-electron-like interface-state band of the
NaCl/Cu(111)~\cite{simon1976bound}. This LIS was observed as a narrow
resonance just below the bottom of the interface-state band in the
differential conductance (d$I$/d$V$) spectra, whereas the VS showed up
as a broad negative ion resonance at a sample voltage of about 2.8\,V
in d$I$/d$V$ spectra. The negative ion resonance was found to have a
large Gaussian broadening (full-width half maximum (FWHM) of $\sigma =
0.27\,\text{V}$) because of the strong electron--phonon coupling of the electron
in the VS to optical phonons in the NaCl bilayer. Despite this strong
coupling corresponding to an estimated Huang--Rhys parameter of about $S = 30$, the
associated relaxation energy of the electron in the VS is not large
enough to prohibit tunneling into unoccupied metal states and to allow
the formation of a stable occupied VS state corresponding to a
localized polaron. In contrast to the VS, the LIS was found to be
dominated by lifetime broadening with a negligible phonon broadening
in the d$I$/d$V$ spectra, owing to the efficient screening of the
electron--phonon interaction by the metal electrons.\\

\begin{figure*}[t]
\includegraphics[width=0.8\textwidth]{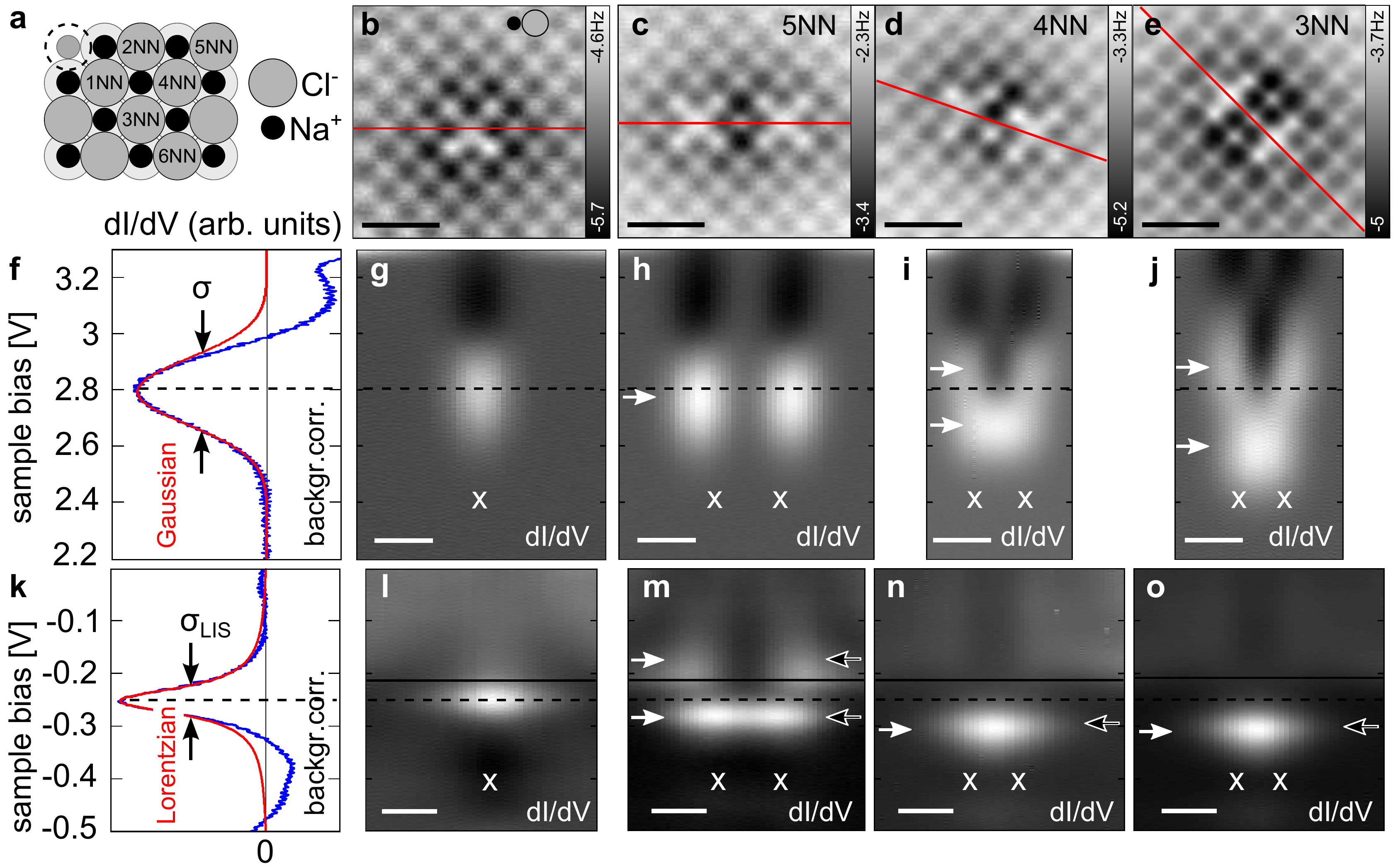}
\caption{\label{fig:dIdV_divac}{\bf Vacancy states and localized interface-states of Cl divacancies.} {\bf a} NaCl bilayer model with a Cl
  vacancy (dashed circle) and its nearest Cl neighbours
  indicated. {\bf b-e} Constant-height CO tip AFM images of a single
  Cl vacancy (b) and divacancies in the 5NN (c), 4NN (d) and 3NN (e)
  configurations. {\bf f} Background-subtracted d$I$/d$V$ spectrum of
  the monovacancy state (VS) peak with a Gaussian fit in red.  {\bf
    g-j} d$I$/d$V(x,V)$ maps along the red lines in b-e (same $V$
  scale as in f). Peaks in d$I$/d$V$ are displayed as bright, whereas
  the darkest areas mark regions of negative differential conductance
  (NDC)\cite{NDCcomment}. The dashed line indicates the monovacancy VS
  peak position. {\bf k} Background-subtracted d$I$/d$V$ spectrum of
  the localized interface-state (LIS) of a monovacancy with a
  Lorentzian fit in red. {\bf l-o} d$I$/d$V(x,V)$ maps along the red
  lines indicated in b-e, respectively (same $V$ scale as in k). The
  dashed line indicates the monovacancy LIS peak position and the
  solid line the interface-state band onset. White crosses
  indicate the vacancy positions. White arrows mark the peak
  positions of the observed d$I$/d$V$ resonances. Black arrows in
  m-o mark the peak positions of the calculated
  LDOS using the $\sigma$-wave multiple scattering model. All STS
  measurements were acquired with Cu tips. Scale bars:
  \unit{10}{\AA}.}
\end{figure*}

The atomically precise locations of intentionally created Cl
divacancies in a NaCl bilayer on Cu(111) were resolved unambiguously
using noncontact atomic force microscopy (AFM) with CO-terminated
tips (CO tips)~\cite{gross2009chemical,mohn2013different}. Divacancies in the
nineth-nearest-neighbour (9NN), 6NN, 5NN, 4NN, 3NN and 2NN
configurations (see schematic in \figref{fig:dIdV_divac}a) were
created from two Cl vacancies close to each other but not in the
vicinity of any other vacancies or defects. Constant-height AFM images
of the 5NN to 3NN divacancies and of a single vacancy are shown in
\figref{fig:dIdV_divac}b-e.
The electronic structure of the VSs and the LISs of the divacancies
were characterized by constant-height d$I$/d$V(x,V)$ maps along the
line connecting the two vacancies in \figref{fig:dIdV_divac}g-j and
\figref{fig:dIdV_divac}l-o, respectively. All d$I$/d$V$ data shown are
recorded using Cu-terminated tips. For the 4NN, 3NN and 2NN~\footnote{
The 2NN divacancy was less stable than the divacancies with larger
separations.}  divacancy configurations, two peaks with Gaussian
line shapes were observed around 2.8\,V. The d$I$/d$V(x,V)$ maps show
that the state at lower energy is localized between the vacancies
(symmetric), whereas the state at higher energy has a nodal plane
between them (antisymmetric), in analogy with the bonding and
antibonding orbitals of the hydrogen molecule. Similarly, the
divacancy LISs were also observed to split into a symmetric and
an antisymmetric state. However, for small vacancy separations, the
latter state lies above the NaCl/Cu(111) interface band onset at
\equnit{$\varepsilon_B$}{-230}{mV} and overlaps with the band
continuum. Hence, only the symmetric state is observed as a
well-defined peak in d$I$/d$V$. The splitting of both the VSs and LISs
was found to increase with decreasing inter-vacancy distance. Tab.~S1
summarizes the observed peak positions and energy splittings of the
VSs and LISs.\\

\begin{figure}
\includegraphics[width=\columnwidth]{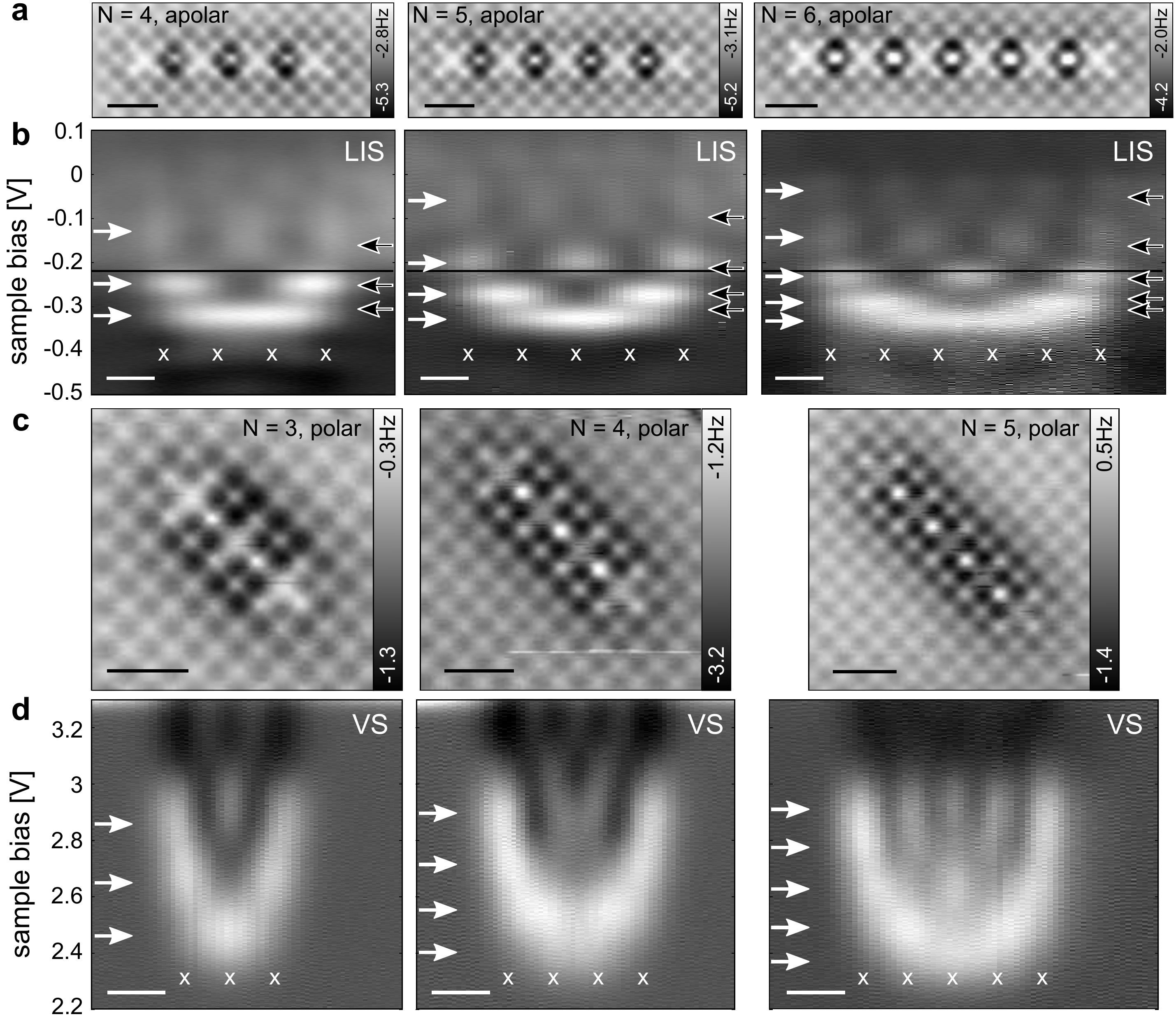}\caption{\label{fig:vac_chains}
  {\bf Vacancy states and localized interface-states of Cl vacancy
    chains.} {\bf a} Constant-height CO tip AFM images of Cl vacancy chains
  with $N = 4,5,6$ vacancies in the 5NN configuration (apolar
  direction). {\bf b} Corresponding d$I$/d$V(x,V)$ maps of
  the LISs along the chains. The continuous line indicates the interface-state band
  onset. White arrows mark the peak positions of the observed
  d$I$/d$V$ resonances, and black arrows mark the peak positions of
  the calculated LDOS. {\bf c} Constant-height CO tip AFM images of Cl vacancy
  chains with $N = 3,4,5$ vacancies in the 3NN configuration (polar
  direction). {\bf d} Corresponding d$I$/d$V(x,V)$ maps of
  the VSs along the chains. White crosses mark the positions of
  the vacancies and the white arrows d$I$/d$V$ peaks. Scale bars: \unit{10}{\AA}.}
\end{figure}

\begin{figure}
\includegraphics[width=\columnwidth]{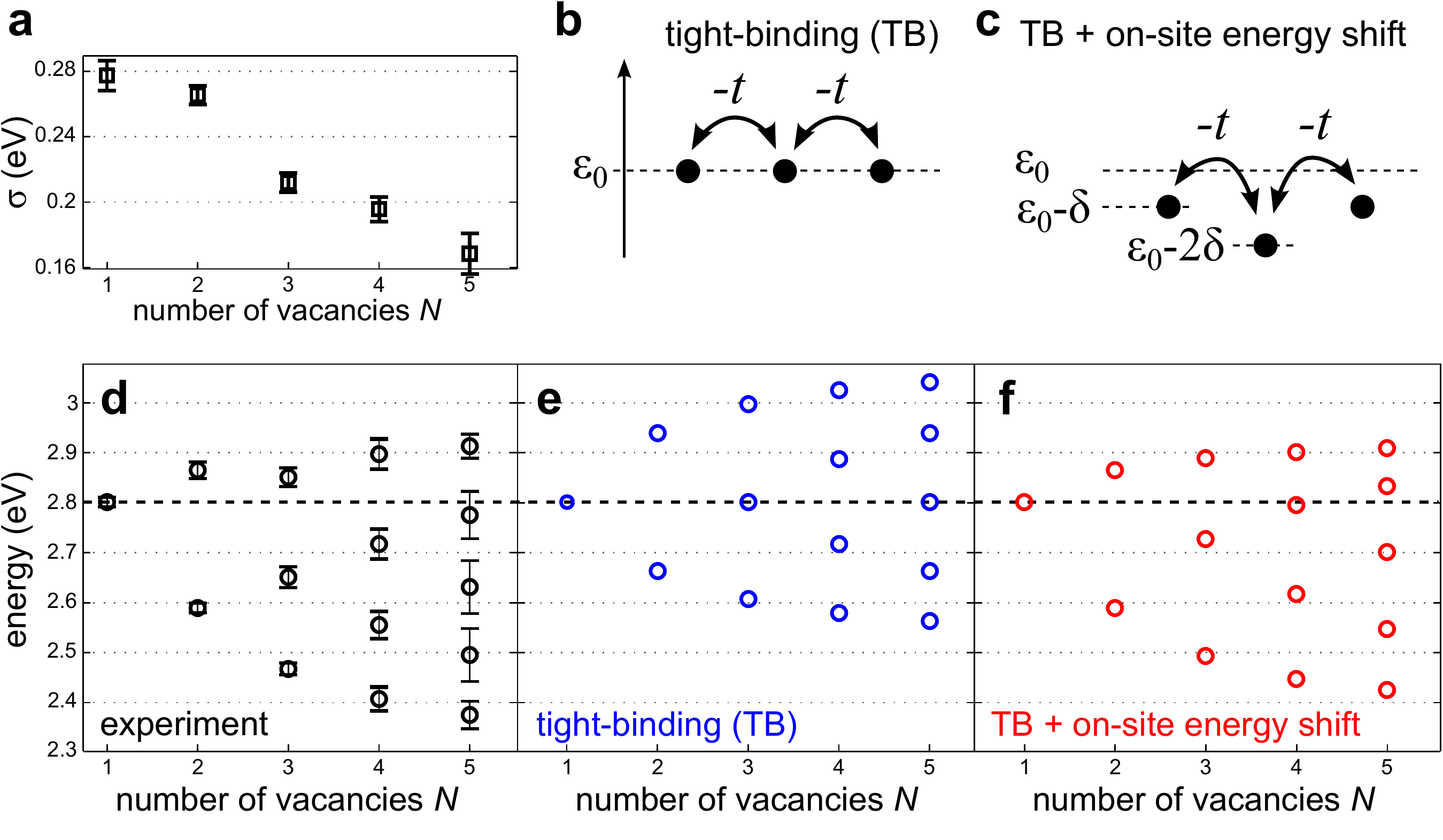}\caption{\label{fig:TBmodel}
  {\bf Tight-binding models.} {\bf a} Experimental VS broadening
  $\sigma$ as a function of the 3NN chain length. {\bf b}
  Tight-binding (TB) model for a $N=3$ chain. {\bf c} TB model
  including the electrostatic on-site energy shift between nearest
  neighbours. {\bf d} Experimental VS energies as a function of the
  3NN chain length. {\bf e} TB model energies obtained by fitting the on-site
  energy $\varepsilon_0 = 2.8\,\text{eV}$ and hopping term
  $t=0.14\,\text{eV}$ to the experimental mono- and divacancy VS
  energies shown in d. {\bf f} TB model accounting for a shift of the
  on-site energy ($\delta=0.07\,\text{eV}$) due to the electrostatic
  interaction between nearest neighbours.}
\end{figure}

To study the delocalization of VSs and LISs further, short one-dimensional arrays
of $N$ vacancies, referred to in the following as vacancy chains, were
created in the 5NN (apolar NaCl direction) and the 3NN (polar NaCl
direction) configuration. They were also characterized by
constant-height CO tip AFM images (\figref{fig:vac_chains}a,c). For
the 5NN chains, the LISs form one-dimensional quantum-well states, observed as
distinct resonances in d$I$/d$V$, with an increasing integer number of
nodal planes with voltage (see \figref{fig:vac_chains}b and
Fig.~S2). Surprisingly, also broad resonances are observed up to
\unit{200-300}{mV} above the interface-state band onset. These
quasi-bound LISs have a lower height in d$I$/d$V$, owing to the
increased lifetime broadening compared to the bound LISs. 
This originates from the additional decay channel into the interface-state band. Furthermore, the
energy position of the ground-state LIS does not change significantly
from the $N = 4$ to the $N=6$ chain and seems to converge to about
\unit{-0.33}{V}. The VSs in the 5NN chains only show a single
resonance peak at about \unit{2.7}{V} because of the weak electronic
interaction between the VSs at these large intervacancy distances
(cf. \figref{fig:dIdV_divac}h). In contrast, the 3NN chains displayed
in \figref{fig:vac_chains}c have a significant interaction between the
VSs and exhibit several resonances, as shown in
\figref{fig:vac_chains}d. The increasing number of nodal planes with
voltage shows that the VSs are delocalized over these chains despite
the strong electron--phonon coupling. In addition, the band width of
VSs increases with chain length $N$, and is more or less saturated at
\unit{0.54}{V} for the $N=5$ chain. Interestingly, the level
broadening decreases for longer chains as displayed in
\figref{fig:TBmodel}a. This finding can be rationalized by the larger
spatial extent of the eigenstates, which decreases their relaxation
energy and accordingly the effective electron--phonon coupling.\\

In the following, the VS energies of the 3NN chains are compared with
the results from a simple tight-binding (TB) model. First, we assume
that each vacancy in the chain has an on-site energy $\varepsilon_0$
and is coupled to its next neighbour(s) with a hopping term $-t$, as
sketched in \figref{fig:TBmodel}b. By adjusting the free fitting
parameters $\varepsilon_0$ and $t$ to the experimental mono- and
divacancy VS energies, one can calculate the energy levels for
different chain lengths as shown in \figref{fig:TBmodel}e. In
comparison with the experimental energy spectrum, depicted in
\figref{fig:TBmodel}d, the level spacings for each chain could be
roughly reproduced, but the TB model does not capture the shift of the
band center towards lower energies with increasing chain length. This
shift can be explained by the electrostatic interaction between
adjacent vacancies that deepens the potential at each vacancy and
therefore lowers the on-site energy. To account for the energy
shift we introduce the parameter $\delta$, which is the on-site energy shift due to a
neighbouring vacancy (see \figref{fig:TBmodel}c). The resulting energy levels of this extended TB
model are shown in \figref{fig:TBmodel}f. The calculated energy
spectrum now agrees very well with the experimentally observed VS levels
of the chains.\\

So far, the strong electron--phonon coupling, as shown by the large
broadening of the VSs, was neglected in our tight-binding models.  The
role of this coupling in the formation of delocalized states by a
tunneling electron in vacancy pairs (and chains) needs to be
investigated.  Here, we use a simple generalization of the
electron--phonon interaction model of the single
vacancy~\cite{Repp2005a} to vacancy pairs and chains.  As detailed in
the SI, the direct electronic interactions between the VSs are
described by a simple tight-binding model, and each VS is coupled
linearly to the phonon modes of the NaCl film.  As we will show next, in this model,
the strong electron--phonon coupling has a profound
effect on the energy splitting and broadening of the vacancy
states. This effect is illustrated here for a divacancy using
appropriate parameters, where each single vacancy state is
coupled to a single Einstein phonon mode. Here, the electronic
interaction between the vacancies is large compared to the
phonon energy, corresponding to an electronically adiabatic
regime.\\

\begin{figure}
\includegraphics[width=\columnwidth]{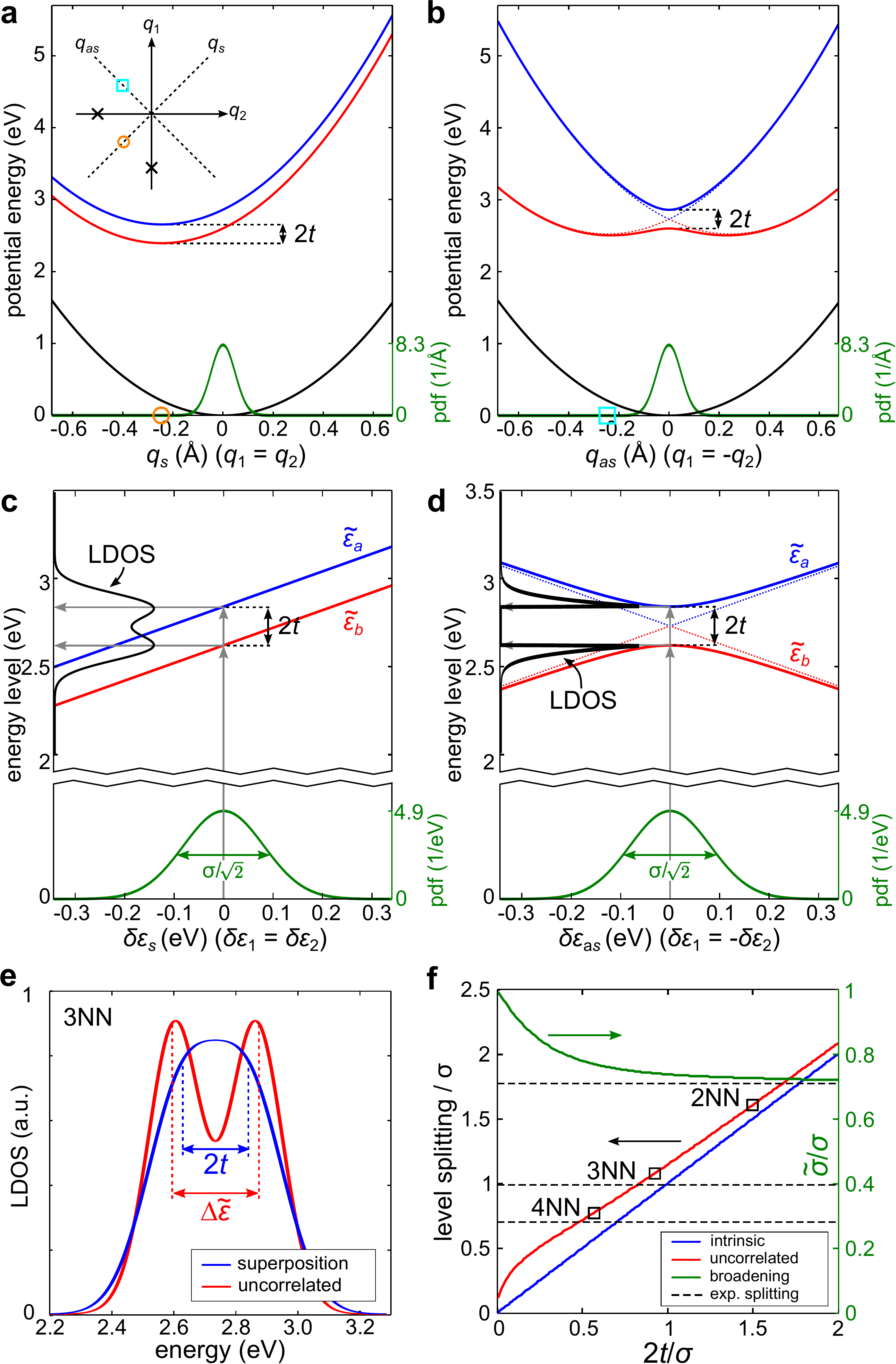}
\caption{\label{fig:split_divac}{\bf Lineshapes of vacancy states of a
    Cl divacancy.}  {\bf a,b} Calculated adiabatic potential energy
  surfaces (PESs) of the electronic ground-state (black) and the
  electronically excited and singly occupied divacancy states (red and
  blue) as a function of symmetric $q_s$ ({\bf a}) and antisymmetric
  $q_{as}$ ({\bf b}) phonon displacements. The vibrational ground-state probability distribution functions (pdf) of $q_s$ ({\bf a}) and $q_{as}$
  ({\bf b}) are indicated by the green curves. Here, $2t$ is the
  intrinsic electronic level splitting. The inset in ({\bf a}) shows
  $q_s$ and $q_{as}$ in terms of the displacements $q_1$ and $q_2$ of
  the two Einstein modes each associated with a vacancy. The black
  crosses mark equilibrium displacements of vacancy 1 and 2 when the
  divacancy is occupied by a single electron. {\bf c,d} Energy-level
  diagram for the divacancy states as a function of symmetric
  $\delta\varepsilon_s$ ({\bf c}) and antisymmetric $\delta\varepsilon_{as}$
  fluctuations ({\bf d}) of the monovacancy energies induced by the phonon
  motion. The vertical gray arrows indicate the electron-attachment
  process and the black curves the resulting LDOS. {\bf e} Calculated LDOS of the 3NN divacancy for uncorrelated vacancy energy fluctuations (red) compared to a superposition of two Gaussians separated by $2t$ and each with a broadening $\sigma$ (blue). $\Delta\tilde{\varepsilon}$ denotes the level
  splitting for uncorrelated vacancy energy fluctuations. {\bf f}
  Scaled level splitting
  $\Delta\tilde{\varepsilon}/\sigma$ as obtained
  from a Gaussian fit to the calculated lineshapes of the divacancy
  states for uncorrelated vacancy energy
  fluctuations (red) compared to the intrinsic splitting (blue) as a function of the scaled intrinsic level
  splitting $2t/\sigma$. The calculated and observed scaled splittings
  for different divacancy configurations are indicated by black
  squares and dashed horizontal lines, respectively. The green curve
  shows the apparent level broadening $\tilde{\sigma}$ for
  uncorrelated vacancy energy fluctuations.}
\end{figure}

In \figref{fig:split_divac}a, the adiabatic potential energy surfaces
(PES) of the singly occupied, divacancy states for symmetric
displacements $q_s$ (where $q_1=q_2$) of the two Einstein phonon modes are
shown. They are split by $2t$, where $-t$ is the direct electronic
interaction between the VSs, and both have the same shape as for the
adiabatic PES of a monovacancy. In contrast, along antisymmetric
displacements $q_{as}$ (where $q_1=-q_2$) (\figref{fig:split_divac}b), the
corresponding PESs exhibit an avoided crossing with a minimum energy
gap of $2t$. This level repulsion will be shown to have a profound
influence on the phonon broadening and the formation of delocalized
states.\\

As for the single vacancy, the result for the phonon broadening of the
tunneling through the divacancy states turns out to have a simple
physical form in the prevailing strong electron--phonon coupling
limit, as detailed in the SI. The phonon broadening is determined by
probabilities for vertical Franck--Condon transitions from the
vibrational ground-state of the PES of the electronic ground-state to
vibrational states on the two PESs for the divacancy states (gray
arrows in \figref{fig:split_divac}c and d). In the strong
electron--phonon coupling limit, the transition probabilities are
dominated by the contribution from the linear electron--phonon
coupling terms. The corresponding phonon broadenings are then simply
obtained from the change in the divacancy energy levels
$\tilde{\varepsilon}_b$ (bonding state) and $\tilde{\varepsilon}_a$
(antibonding state) with the Gaussian fluctuations of the energies
$\delta\varepsilon_1$ and $\delta\varepsilon_2$ of the vacancy 1 and 2
from their mean value $\varepsilon_0$ due to the zero-point motions of
the phonons. As shown in the SI, this result is also valid for more
general phonon baths and electron--phonon couplings. \\

The strong coupling result for the phonon broadening of the divacancy
states is illustrated in \figref{fig:split_divac}c and d. A symmetric
fluctuation $\delta\varepsilon_s$ (where $\delta\varepsilon_1 =
\delta\varepsilon_2$) results, as shown in \figref{fig:split_divac}c,
in a linear dependence of $\tilde{\varepsilon}_b$ and
$\tilde{\varepsilon}_a$ on $\delta\varepsilon_s$ with slope one, but they are split by
$2t$. These Gaussian fluctuations then give rise to two overlapping
Gaussian lineshapes centered at $\varepsilon_0\pm t$ with a variance
of $\left\langle\delta\varepsilon_s^2\right\rangle$
for the local density of states (LDOS).\\

In contrast,
the level repulsion results in a minimum energy separation $2t$
between $\tilde{\varepsilon}_b$ and $\tilde{\varepsilon}_a$ for an
antisymmetric fluctuation $\delta\varepsilon_{as}$ (where $\delta\varepsilon_1 =
-\delta\varepsilon_2$). Furthermore,
$\delta\varepsilon_{as}$ breaks the symmetric and antisymmetric
character of the two coupled VSs, and they tend to localize on each
vacancy for $\delta\varepsilon_{as} \gg t$. The resulting LDOS from
these Gaussian fluctuations, shown in \figref{fig:split_divac}d, consists of two relatively sharp, highly asymmetric lineshapes, with mean energies separated by more
than the energy splitting $2t$.  In general, the fluctuations
$\delta\varepsilon_1$ and $\delta\varepsilon_2$ of the two vacancy
energies will be a superposition of both $\delta\varepsilon_{s}$ and
$\delta\varepsilon_{as}$. In the case of uncorrelated energy
fluctuations
$\left\langle\delta\varepsilon_1\delta\varepsilon_2\right\rangle=0$,
the LDOS is a convolution of the lineshape from
$\delta\varepsilon_{as}$ with a Gaussian lineshape with a variance
of $\left\langle\delta\varepsilon_{1,2}^2\right\rangle/2$
because $\delta\varepsilon_{s}$ and
$\delta\varepsilon_{as}$ contribute equally to 
$\left\langle\delta\varepsilon_{1,2}^2\right\rangle$,
the variance of the single vacancy energy fluctuations.
As the Gaussian shape dominates the broadening in such a convolution,
the resulting LDOS as displayed in \figref{fig:split_divac}e 
has a similar shape as two Gaussians with a FWHM reduced by almost $1/\sqrt{2}$
as compared to an isolated vacancy.
This decrease of
the broadening in the uncorrelated case is the underlying reason why the observed broadenings of the
levels of the chains decrease with chain size (see SI).\\

The two phonon modes considered here affect only one of the VSs each
and result in uncorrelated vacancy site energy fluctuations. 
In reality, however, a given individual phonon mode may act on the energy of 
both vacancies in a symmetric or antisymmetric fashion
(consider e.g. the motion of an ion centered between two vacancies). 
Hence, partially correlated vacancy site energy fluctuations cannot be excluded.
As shown in the SI, 
the two Einstein modes and their symmetric and antisymmetric fluctuations
can serve as a basis for the description of any phonon bath and
electron--phonon coupling.
Our experimental results are consistent with the assumption of fully uncorrelated fluctuations
of the two VSs.
Interestingly, in the uncorrelated case, the LDOS,
shown in \figref{fig:split_divac}e (red line), exhibits two peaks with an
apparent separation that is substantially larger than the intrinsic
splitting of $2t$. This result is a characteristic effect of the level
repulsion between $\tilde{\varepsilon}_b$ and $\tilde{\varepsilon}_a$
and is in sharp contrast compared to a superposition of two monovacancy resonances separated by the intrinsic splitting (shown in \figref{fig:split_divac}e (blue line)). In the latter case, one would not be able to identify separate peaks due to the larger broadening and smaller splitting.\\

The enhancement of the apparent peak
separation in the uncorrelated case over the intrinsic splitting is shown
in \figref{fig:split_divac}f as a function of the scaled intrinsic
energy splitting $2t/\sigma$, where $\sigma$ is the broadening of 
a single VS. We define the apparent peak separation
$\Delta\tilde{\varepsilon}$ as the separation between the centers of
the two Gaussian lineshapes with broadening (FWHM) $\tilde{\sigma}$
fitted to the calculated LDOS. The
relative enhancement $\Delta\tilde{\varepsilon}/2t$ of the apparent
splitting is most important for small $t$. 
For large $t$, the splitting of divacancy energy levels approaches the 
intrinsic energy splitting of $2t$. 
As for the single vacancy level, the broadening of a divacancy level then scales with the square root of its
relaxation energy (see SI).\\

The hopping terms $t$ can be extracted from \figref{fig:split_divac}f
by taking $\Delta\tilde{\varepsilon}$ equal to the observed splitting
and using the single vacancy value 0.27 eV for $\sigma$. For example,
$t =0.11$ eV was obtained from $\Delta\tilde{\varepsilon}=0.28$ eV
for the 3NN divacancy. 
The values obtained for the hopping terms
$t$ are very reasonable as supported by density-functional theory (DFT) calculations of the unoccupied
Kohn--Sham states of Cl divacancies in NaCl(2ML)/Cu(111). As shown in
\figref{fig:split_divac}f, the calculated values for $2t$ of the
various divacancy configurations are close to the
values for the intrinsic splittings extracted from experimental data. However,
the absolute divacancy energies are underestimated by the DFT
calculations, as expected (see Fig.~S3e).\\

The coupling between the LISs of Cl divacancies and chains were
studied in a simple, multiple $\sigma$-wave scattering model of
the interface-state scattering from the positively charged vacancies
(see methods and SI). The results from this model, as indicated in
\figref{fig:dIdV_divac}l-o, are able to fully capture the observed
behavior of the coupled LISs. The absence of a LIS doublet below the
onset of the interface-state band is simply understood by the
relatively strong interaction between the weakly bound LISs of the two
vacancies at small intervacancy distances. With increasing distances
between the vacancies, this interaction decreases, and a second LIS
appears below the band onset.  The observed symmetric character of the
LIS below the band edge and antisymmetric character of the extended
LIS above $\varepsilon_B$ for the 3NN vacancy are also revealed in the
calculated LDOS images (see Fig.~S5). Furthermore, the obtained energies
for the LISs of the vacancy chains are in excellent agreement with the
experiments, as indicated in \figref{fig:vac_chains}b. \\

In summary, from an AFM and STS study we find that the localized VSs and LISs at
intentionally created Cl divacancies in a NaCl bilayer on Cu(111) form
symmetric and antisymmetric states, in analogy to the bonding and
antibonding orbitals of a hydrogen molecule. As expected, the energy
splitting between these states increases with decreasing intervacancy
distance. 
A comparison with theory shows that the energy splitting of the VSs is significantly enlarged by the strong coupling of the tunneling
electrons with phonons in the NaCl film. 
Furthermore, VSs and LISs of vacancy chains form one-dimensional quantum-well states. The VS levels of the
chains could be well described by a simple tight-binding model that
takes the electrostatic interaction between neighbouring vacancies
into account. 
The model provides also a microscopic understanding why the level broadening is reduced with increasing chain size.
Already for about five coupled vacancies, a one-dimensional electronic band structure evolves.

\section*{Methods}
\subsection*{Sample preparation and Cl vacancy creation.}
The experiments were carried out in a home-built low-temperature
combined STM and AFM operated at 5\,K. A Cu(111) single-crystal sample
was cleaned by several sputtering and annealing cycles. NaCl was
evaporated thermally, keeping the sample temperature at about 270\,K,
such that defect-free, (100)-terminated NaCl bilayer islands were
formed~\cite{bennewitz1999ultrathin,repp2004snell}. Bias voltages
refer to the sample voltage with respect to the tip. In the
spectroscopic measurements, the tunneling conductance d$I$/d$V$ was
recorded with conventional lock-in techniques with an ac bias
amplitude of 25\,mV at a frequency of 294\,Hz. In the double-barrier
tunneling junction geometry, the voltage drop across the insulating
film will cause a voltage-dependent shift of the electronic
levels. This tip-induced Stark shift is a few percent of the applied
bias but depends only weakly on the tip distance. Therefore it is not
considered here~\cite{Repp2005a}.  The vacancies were created by
bringing the Cu tip into controlled contact with the NaCl
surface. Thereby a Cl atom is transfered to the tip apex, as evidenced by a
characteristic contrast change in STM images and a remaining depression
at the predefined Cl site. Constant-height AFM images with a CO tip~\cite{gross2009chemical,mohn2013different} were
recorded to identify the vacancy location and exclude the generation of other close-by
defects. In addition, we measured the
local contact potential difference on the polar film with Kelvin probe
force microscopy to ensure that the defects are indeed Cl
vacancies~\cite{Grossvac2013}. 

\subsection*{Theoretical calculations.}
\paragraph{Density functional calculations.}
The electronic and geometric structure of the divacancies and vacancy
chains in a NaCl bilayer on a Cu(111) surface were obtained using
periodic density functional theory (DFT) calculations. We used the
projector augmented wave (PAW)
method~\cite{kresse1999ultrasoft,blochl1994projector} as implemented
in {\tt VASP}~\cite{Kresse1996}. The exchange-correlation energy was
described by the optB86b+vdW approximation\cite{PhysRevB.76.125112,Klimes2011}. The
incommensurate growth of NaCl on Cu(111) was modeled by a super cell
consisting of 90 Cu atoms and 32 Na and Cl atoms in each layer (4
layers of Cu and 2 layers of NaCl), which corresponds to about 2\%
mismatch of NaCl distances compared to those in bulk NaCl.
\paragraph{Phonon broadening model.}
This simple model detailed in the SI contains two key
parameters: the phonon broadening $\sigma$ (FWHM) of the single VS and the magnitude of the direct
interaction energy $-t$ between the two VSs. The value for $\sigma$ is
determined from the observed broadening of 0.27\,eV of the single
vacancy VS. The values for $t$ were adjusted to reproduce the observed
energy splittings of the divacancy VSs and compared with results from the DFT calculations. For further information see the SI~\cite{suppinfo}.
\paragraph{Interface-state band scattering model.}
The coupling between the LISs of Cl divacancies are modeled with a
multiple $\sigma$-wave scattering model of the interface-state
scattering from the vacancies. This model is detailed in the
SI~\cite{suppinfo}.  The interface-state band is modeled by a 2D
free-electron band with an effective mass $m^*$ and a band onset
$\varepsilon_B$.  The values for $m^*= 0.46\,m_e$ and $\varepsilon_B =
-230\,\text{meV}$ of the interface-state band were taken from STM and
STS measurements\cite{repp2004snell}. The remaining parameter
$\varepsilon_0 - \varepsilon_B$, determines the $t$-matrix element
for $\sigma$-wave scattering and was fitted to the observed value of -19 meV for the
LIS of the vacancy.

\section*{Acknowledgments}
We thank R.~Allenspach for comments and acknowledge financial support
from the ERC Advanced Grant CEMAS (agreement no. 291194) and the EU projects PAMS (610446) and
QTea (317485). Allocation of computer resources through SNAC and CSC is gratefully
acknowledged.\\

\section*{Author contributions}
All authors contributed to the preparation of the paper. B.S., N.P.,
L.G, G.M. and J.R. conducted the experiments. M.P. and S.P. developed
the coupling models and performed the DFT calculations.

\section*{Additional information}
Correspondence and requests for materials should be addressed to B.S. (e-mail: bsc@zurich.ibm.com).

\clearpage

\end{document}